# MULTIPLE-CHARGE BEAM DYNAMICS IN AN ION LINAC

P.N. Ostroumov, J.A. Nolen, K.W. Shepard
Physics Division, Argonne National Laboratory,
9700 S. Cass Avenue, Argonne, IL, 60439

*Abstract*

There is demand for the construction of a medium-energy ion linear accelerator based on superconducting rf (SRF) technology. It must be capable of producing several hundred kilowatts of CW beams ranging from protons to uranium. A considerable amount of power is required in order to generate intense beams of rare isotopes for subsequent acceleration. At present, however, the beam power available for the heavier ions would be limited by ion source performance. To overcome this limit, we have studied the possibility of accelerating multiple-charge-state (multi-Q) beams through a linac. We show that such operation is made feasible by the large transverse and longitudinal acceptance which can be obtained in a linac using superconducting cavities. Multi-Q operation provides not only a substantial increase in beam current, but also enables the use of two strippers, thus reducing the size of linac required. Since the superconducting (SC) linac operates in CW mode, space charge effects are essentially eliminated except in the ECR/RFQ region. Therefore an effective emittance growth due to the multi-charge beam acceleration can be minimized.

## 1 INTRODUCTION

A preliminary design and beam dynamics study has been performed for the rare isotope accelerator (RIA) driver linac structure and is discussed elsewhere [1,2]. A schematic view of the linac is shown in Fig.1.

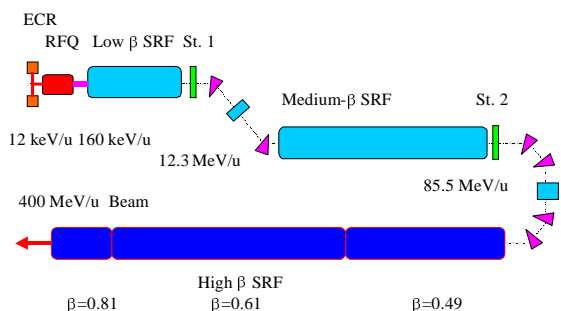

Figure 1: Simplified layout of the Driver Linac.

The linac contains three main sections: a "pre-stripper" section up to the first stripping target at 12.3 MeV/u, a medium energy section defined and separated by the stripper targets and a high energy section with a maximum uranium energy of 400 MeV/u. Total voltage of the linac is 1.36 GV. The pre-stripper section consists of an ECR ion source followed by mass and charge selection, an initial linac section consisting of an RFQ and 96 low-beta independently-phased SRF cavities. The middle section is based on 168 intermediate-beta SRF cavities. The high-energy section consists of 172 elliptical cavities designed for three different velocities. The heaviest ions, which are not fully stripped at the first stripper, will be stripped a second time at ~85 MeV/u. The charge state distribution of uranium ions is centered at the charge state $q_0$=+75 at 12.3 MeV/u from the first stripper. Five charges encompassing ~80% of the incident beam after the first stripper will be accelerated simultaneously in the medium-β section. After the second stripper, 98% of the beam is in five charge states neighbouring $q_0$ = 89, all of which can be accelerated to the end of the linac.

The accelerating field taking into account the cryostat filling varies from 1.6 MeV/m to 5 MeV/m. Transverse beam focusing over all of the driver linac is provided by SC solenoids. The length of the focusing period depends on the resonator type.

The behaviour of the uranium multi-Q beam has been studied both by analytical and numerical methods. The effects of various factors, such as beam mismatch, misalignments, accelerating field errors and other factors affecting the emittance growth of a multi-Q beam are discussed.

## 2 BEAM DYNAMICS

### 2.1 Longitudinal beam dynamics

When a particle with a charge state, q, and mass number, A, traverses an accelerating cavity of length, $L_c$, and electric field $E=E_g(z)cos\omega t$, the energy gain per nucleon $\Delta W_{s,n}$ is determined by the expression $\Delta W_{s,n} = \frac{q}{A} eE_0 T(\beta,\beta_G) L_c \cos\varphi_s$, where $T(\beta,\beta_G)$ is the transit time factor, $E_0$ is the average accelerating field of the cavity and $\varphi_s$ is the synchronous phase. $\beta_G$ is the geometrical beta of the cavity. The transit time factor (TTF) is a complicated function of both the field distribution and the particle velocity. At low energy, the particle velocity may change appreciably during the passage through a multiple gap cavity. For this reason, the TTF is most conveniently calculated numerically.

We define the synchronous phase for a given particle traversing a given field with respect to that rf phase producing maximum energy gain. The synchronous phase, as with the TTF, is generally most conveniently determined numerically. The synchronous motion of an ion with charge state $q$ can be considered as motion in an equivalent traveling wave with the amplitude $E_m = E_0 T(\beta, \beta_G)$. For beam energies higher than several MeV/u the accelerating field $E_m$ can be considered as changing adiabatically along the linac. At lower energies the adiabatic conditions of ion motion are not valid due to the large increment of beam velocity in the cavity.

A heavy-ion linac is usually designed for the acceleration of many ion species. In a SC linac the cavities, fed by individual rf power sources, can be independently phased. The phase setting can be changed to vary the velocity profile for synchronous motion along the linac. For a given, fixed phase setting, the synchronous velocity profile, and the TTF profile are fixed along the accelerator. To accelerate ions with a charge-to-mass ratio $(q/A)_i$ different from the design value, the following relation must be satisfied $\left(\frac{q}{A}\right)_i E_{m,i} \cos\varphi_{s,q_i} = \left(\frac{q}{A}\right)_0 E_{m,0} \cos\varphi_{s,q_0}$. Thus, the velocity and the accelerated beam energy per nucleon do not depend on the ion species.

In an independently phased cavity array such as an SRF ion linac, beams of different charge-to-mass ratio can be accommodated by changing either or both the phase and amplitude of the electric field. Allowing both parameters to vary permits the option of varying the velocity profile. This can provide higher energies per nucleon for ions with a higher charge-to-mass ratio.

The RIA driver linac will accelerate uranium ions at charge state $q_0=75$ after the first stripper and at $q_0=89$ after the second stripper. The simultaneous acceleration of neighbouring charge states becomes possible because the high charge-to-mass ratio makes the required phase offsets small for a limited states of charge states. We note that different charge states of equal mass will have the same synchronous velocity profile along the linac if the condition

$$\left(\frac{q}{A}\right)_i \cos\varphi_{s,q} = \left(\frac{q}{A}\right)_0 \cos\varphi_{s,0} \quad (1)$$

is fulfilled. The simultaneous acceleration of ions with different charge states requires an injection of the beam with each charge state $q$ at a synchronous phase which is determined from (1) $\varphi_{s,q} = -Arc\cos\left[\frac{q_0}{q}\cos\varphi_{s,0}\right]$.

Figure 2 shows the synchronous phase as a function of charge states when the synchronous phase for $q_0=75$ is $\varphi_{s,0} = -30°$. This particular example shows that if the linac phase is set for charge state $q_0=75$, a wide range of charge states can be accelerated. As seen even for charge state 70 only a small change from 30 to 23° in synchronous phase is required. The longitudinal focussing of the linac being considered is sufficient to accept the predicted beam emittance.

The separatrices for charge states $q=73, 75$ and $77$ calculated in a conservative approximation are shown in Fig. 3. Each particle with different charge state q oscillates around its own synchronous phase with slightly different amplitude [2]. It would be entirely feasible to eliminate the relative oscillations. If the linac has been tuned for the acceleration of some charge state $q_0$, then the particle bunches of different, neighbouring charge states could be injected into the linac at different, neighbouring rf phases in order for each charge state to be matched precisely to its own phase trajectory. The higher the charge state, the sooner it must arrive at a given point to be matched. One possible method of adjusting the phase of multiple charge states would be a magnetic system combined with rf cavities.

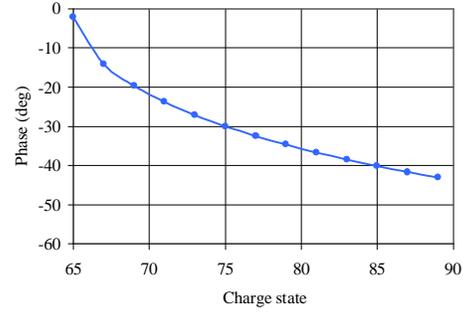

Figure 2: Synchronous phase as a function of uranium ion charge state. $\varphi_{s,0}=-30°$ for $q_0=75$.

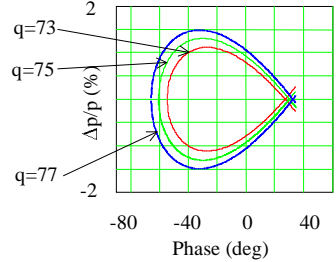

Figure 3: Separatrices in the longitudinal phase space for charge states 73, 75 and 77 of uranium beam.

For most applications, however, such a system is not necessary since the acceleration of a multi-Q beam is possible even without matching the different charge states to their proper synchronous phase. If all charge states are injected at the same time (at the same rf phase), then, as described above, each charge state bunch will perform coherent oscillations with respect to the tuned charge state $q_0$. One can view this as an increase in the total (effective) longitudinal emittance of the multi-Q beam, relative to the (partial) longitudinal emittance of the individual charge state bunches. For the heavy-ion SRF linac being considered, the longitudinal emittance is predominantly determined by the injector RFQ, and can be made as small as *~2.0 keV/u·nsec* for a single charge-state beam [3]. For comparison, the linac acceptance, given by the area of the separatrix shown in Fig. 3, for $q_o=75$ is ~77 keV/u·nsec. As will be shown below, this provides ample headroom for the effective emittance growth introduced by the acceleration of multiple charge states.

It should be noted that if no phase-matching is done for different charge states, additional emittance growth will occur at frequency transitions in the linac. Heavy-ion linacs typically have several such transitions to permit efficient operation over the large velocity range required. In the RIA driver linac the strippers will be placed in the region of frequency transitions thereby eliminating effective emittance growth of multi-Q beams.

## 2.2 Transverse beam dynamics

We now consider the transverse phase space for this same uranium beam through the medium-β section of the linac. The focussing period is defined by a lattice of a SC solenoid following each pair of SRF cavities. The present linac design calls for solenoid focussing elements because SC solenoids are cost effective for this application, but the following analysis is not particularly restricted by this choice.

Table I shows the Twiss parameters for 5 different charge states calculated for the focusing period at the 12 MeV/u region of the linac. The difference in Twiss parameters for five charge states is sufficiently small that all the charge states can be injected into the linac with the same transverse parameters producing an effective emittance growth of *6.5%*.

The transverse beam emittance is determined by the ECR source. Present-day ECR sources can produce beam intensities up to ~1 pμA for a single charge state of uranium, with an normalized emittance (containing 90% of the particles) equal to ~0.2 π·mm·mrad [4]. We compare this emittance with the transverse acceptance of the solenoidal focussing channel of the driver linac assuming $\mu_x=60°$, $\beta_{x,max} = 3.17$ mm/mrad. The maximum value of the $\beta_x$-function occurs at the center of solenoids, which has a bore radius of 15 mm. This implies a normalized acceptance $A_n = 11.6$ π·mm·mrad. The acceptance of this section is ~50 times larger than the beam emittance at the entrance. It should be noted that the acceptance of the next linac section, the high energy part, is even larger, ~100 π·mm·mrad.

Transverse emittance can grow due to several error effects. We discuss below two type of errors which can most seriously impact the multi-Q beam dynamics. The first type of error, mismatch, is caused by errors in tuning or matching the beam into the linac and arises because of errors in measurement of the input beam parameters. The second type of error is transverse displacements of the focussing lenses.

For a low-intensity single charge state beam, mismatched betatron motion and coherent transverse oscillations will not increase effective transverse emittance. For this case, the errors discussed above are often of little consequence and can be easily corrected. In the case of a multi-Q beam, however, the different charge states have different betatron frequencies. As the beam proceeds along the linac, the transverse oscillations of the various charge states eventually become uncorrelated and the effective total emittance, summed over all charge states, increases.

Table I: Twiss parameters of the matched beam after the first stripper for 5 different charge states.

| q | $\alpha_x$ | $\beta_x$ | $\gamma_x$ |
|---|---|---|---|
| 73 | 0.428 | 1.536 | 0.770 |
| 74 | 0.435 | 1.518 | 0.783 |
| 75 | 0.441 | 1.500 | 0.783 |
| 76 | 0.448 | 1.483 | 0.809 |
| 77 | 0.455 | 1.467 | 0.823 |

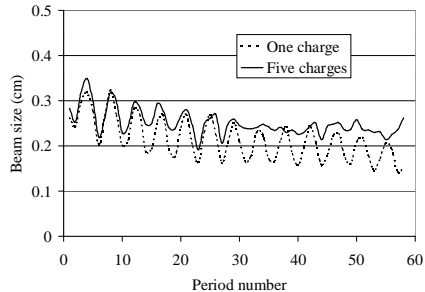

Figure 4: Transverse beam envelopes for one charge state beam (dots) and 5 charge state beam (solid line). Input beam is mismatched by a factor 1.4.

One aspect of this behaviour can be illustrated by considering a mismatched beam through the 58 focussing periods of the linac between the two strippers. While the actual linac lattice will be slightly more complex, it is sufficient for us to consider the periodic focussing structure as having constant length. We assume the solenoids to be tuned for a phase advance over one period of $\mu_x = 60°$, for charge state 75. Although the phase advance per period does not depend strongly on the charge state, over 58 periods the phase differences between different charge states become appreciable. If the input beam is mismatched, the phase space ellipse begins to rotate, at twice the betatron frequency, tracing out a (matched) ellipse of larger area. Fig. 4 shows beam envelopes both for a single charge state beam and also for a five charge state beam. The oscillations of the mismatched beam remain coherent for the single charge state, but not for the multi-Q case. To summarize, the main difference between one- and multi-Q beams is that mismatch of a single charge state beam is generally correctable, and does not lead to transverse emittance growth. For multiple charge states, correction is more difficult, and will generally induce growth in the transverse emittance.

Multi-Q beams are also more severely affected by misalignment errors. Misalignments produce a transverse magnetic field on the linac axis and coherently deflect the beam. For a single charge state beam, misalignment causes lateral displacement of the beam, but no emittance growth so long as the beam remains in the linear region of the focussing elements. With a beam containing multiple charge states, the differing betatron periods, as well as the differing displacements, cause growth in the transverse emittance. We have performed Monte Carlo simulations of the dynamics of multi-Q beams in the presence of alignment errors. We considered a five charge-state uranium beam in that portion of the linac between the first and second strippers. To make the simulation more realistic, we assumed a mismatch factor of 1.2 for the beam out of the first stripper. We introduced alignment errors by displacing separately both ends of each of the 58 focussing solenoids in both x and y by an amount randomly varying over the range ±300 μm. The bar graph in Fig. 5 is a histogram of the simulation results. Note that for some sets of alignment errors, the emittance growth factor can be as high as 8.5.

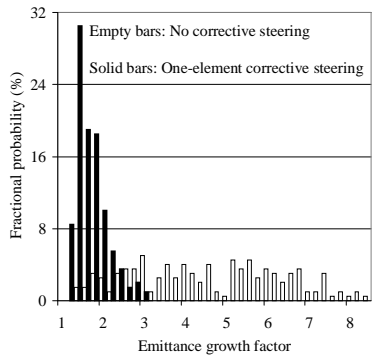

Figure 5: Probability of emittance growth in the misaligned focussing channel. The emittance growth factor is the ratio of (normalized) transverse emittance of the beam at exit to that at entrance.

Even for the multi-Q beams, however, emittance growth can be substantially reduced by simple corrective steering procedures. We have modelled this by assuming a measurement of the beam centroid position and corrective steering to be performed once every four focussing periods. This interval would correspond to the space between cryomodules in the benchmark linac design.

The transverse tune has important effects. We consider the case of a 60° phase advance per focussing period. For this case, the phase advance between the points at which monitoring and steering is performed is $\Phi_x = 240°$. This rotation in transverse phase space transforms a centroid displacement at one corrective station to a deflection at the succeeding station, which can be directly corrected by simple one-element steering at that point. In our simulations, the beam center was calculated as the center of gravity of a five charge state beam with q=73...77. The results are shown in Fig. 5. It can be seen that the emittance growth factor has, in all cases, been reduced to less than 3. For the entire set of cases of random errors, the most probable value for the effective emittance growth factor is 1.4. The steering procedure effectively reduces the divergence in transverse phase space thereby reducing the possible effective emittance growth.

## 3 EXAMPLES OF MULTI-Q BEAM ACCELERTATION

### 3.1 Two charge state beam in the prestripper linac

We have shown [3] that it is possible to accept two charge states from an ECR and accelerate both to the first stripping target at 12 MeV/u. The ECR is followed by an achromatic bend with a charge selector which transports a two-charge-state uranium beam to the entrance of a multi-harmonic buncher upstream of the RFQ. The fundamental frequency of both bunchers is one half of the RFQ frequency. A combination of the multi-harmonic buncher and RFQ bunches more than 80% of each charge state to an extremely low longitudinal emittance (total emittance is lower than ~2.0 $\pi$·keV/u·nsec) beam at the output of the RFQ. The second buncher is located directly before the RFQ entrance and changes the average velocities of each charge state to the design input velocity of the RFQ. Over the distance between the first, multi-harmonic buncher, and second buncher the bunched beams of each charge state are formed with a separation by 360° at the RFQ frequency due to the different average velocities of each charge state. Electrostatic quadrupoles provide focussing and transverse matching to the RFQ acceptance. The results of the design and beam dynamics studies are presented at this conference (see ref. [3]).

### 3.2 Five charge state beam in the medium-β linac

We have carried out a Monte Carlo simulation of multi-Q beam acceleration from the first stripper through the second stripper and continuing to the end of the linac. The simulation starts with a 12.3 MeV/u uranium beam equally distributed over 5 charge states, all at the same rf phase, and with a longitudinal emittance ~1.0 $\pi$·keV/u·nsec. We consider in detail the behaviour of this beam between the two strippers, a section of linac consisting of 3-gap SRF cavities operating at 172.5 MHz and 345 MHz. The rf phase throughout this section has been set for acceleration of uranium with charge state $q_0=75$ at synchronous phase $\varphi_{s,75}=-30°$. The phase is calculated using values of the electric field numerically generated using realistic cavity geometries. The beam tracking simulation was done with a modified version of the LANA code [2]. This code completely simulates beam dynamics in the six-dimensional phase space including alignment errors. The phase space plot of all 5 charge state bunches at $W_n$=85.5 MeV/u, just before the second stripper, is shown in Fig. 6 together with the acceptance of 805 MHz elliptical SRF structure. The acceptance is obtained by Monte Carlo simulation. After the second stripper, the effective longitudinal emittance of the multi-Q beam is increased by a factor of ~6. We note, however, that this longitudinal emittance is still substantially less than the acceptance of the remaining portion of the SC linac.

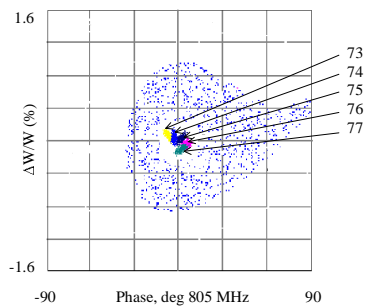

Figure 6: Phase space plots of the multi-Q beam at the location of the second stripper. Large dotted area represents the acceptance of high-energy section.

The effect of rf field errors on longitudinal beam dynamics in a multi-cavity linac becomes significant in the present case mainly because of the large number of individual cavities. These errors are caused by fluctuations, which we assume to be random, in the rf phase and amplitude of the electromagnetic fields in the cavities. We have performed numerical simulation to estimate the effects of this latter class of error, for both single-charge-state and also for multi-Q beams. The error effects are included by introducing phase and amplitude errors for each of the cavities, randomly distributed over the indicated range. We have found [2] that even including both these effects, the total increase in longitudinal emittance is still well below the acceptance of the high energy part of the driver linac, ~77 $\pi \cdot keV/u \cdot nsec$.

The construction of a high-intensity heavy-ion linac requires at least two stripping foils. In order to avoid beam losses in the high-energy part of the linac the low-intensity unwanted charge states must be carefully separated and dumped. It will require a system containing dipole magnets and a rebuncher in order to provide a unit transformation of the 6-dimensional beam phase space (see Fig.1). We have designed such systems for several cases: a 180° bend, a parallel translation of ~4.5 m and a chicane-like system for the straight line transformation. After the stripping target, all charge states have the same velocity, therefore, such a matching system is isopath for different charge states.

### 3.3 Multi-Q beam test at ATLAS

A test of the acceleration of multi-Q beams was performed at the ATLAS accelerator. A $^{238}U^{+26}$ beam from an ECR ion source was accelerated to 286 MeV (~1.2 MeV/u) and stripped just before the 'Booster' section of ATLAS. All charge states near $q_0$=38 were simultaneously accelerated in the Booster. The parameters of each selected charge state were carefully measured. Tuning of the focusing fields to get 100% transmission was accomplished with the $^{58}Ni^{+9}$ guide beam prior to switching to the uranium mixed beam. About 94% transmission of the multi-Q uranium beam was detected. Six charge states were accelerated through the Booster with an average energy spread within 1.5%. Detailed experimental results are given in ref. [5].

## 4 CONCLUSIONS

The large longitudinal and transverse acceptance characteristic of superconducting heavy-ion linacs makes possible the acceleration of multiple charge state beams. Our studies indicate that it is quite feasible to accelerate 2 charge states of uranium from an ECR to the first stripper, 5 charge states of the same beams after the first stripper and 5 charge states after the second stripper in this linac. Such operation could provide ~120 kW of uranium beam using a demonstrated performance of an ECR ion sources.

## 5 ACKNOWLEDGEMENTS

The authors wish to thank our collaborators from several laboratories. R. Pardo, M. Portillo (ANL), V.N. Aseev (INR, Moscow), A.A. Kolomiets (ITEP, Moscow), J. Staples (LBNL) participated in developing of several aspects of multi-Q beam studies.

Work supported by the U. S. Department of Energy under contract W-31-109-ENG-38.